\documentclass[%
superscriptaddress,
 preprint,
 aps,
 pra,
 amsmath,amssymb
 ]{revtex4-1}

\usepackage[english]{babel}
\usepackage[utf8]{inputenc}

\usepackage{amsthm}
\usepackage{mathtools}
\usepackage{physics}
\usepackage{xcolor}
\usepackage{graphicx}
\usepackage{adjustbox}
\usepackage{placeins}
\usepackage[T1]{fontenc}
\usepackage{lipsum}
\usepackage{csquotes}

\usepackage{siunitx}

\usepackage{xr-hyper}

\usepackage{xcolor}
\usepackage[normalem]{ulem}


\usepackage[pdftex, pdftitle={Article}, pdfauthor={Author}]{hyperref} 

\newcommand{\0}{\phantom{0}}

\begin{document}
\title{
MODNet - accurate and interpretable property predictions for limited materials datasets by feature selection and joint-learning
}

\author{Pierre-Paul De Breuck}
    \affiliation{UCLouvain, Institute of Condensed Matter and Nanosciences (IMCN), Chemin des \'Etoiles~8, B-1348 Louvain-la-Neuve, Belgium}
    
\author{Geoffroy Hautier}
\affiliation{UCLouvain, Institute of Condensed Matter and Nanosciences (IMCN), Chemin des \'Etoiles~8, B-1348 Louvain-la-Neuve, Belgium}
\affiliation{Thayer School of Engineering, Dartmouth College, Hanover, New Hampshire 03755, USA}
    
\author{Gian-Marco Rignanese}
	\email{Correspondence to gian-marco.rignanese@uclouvain.be}
	\affiliation{UCLouvain, Institute of Condensed Matter and Nanosciences (IMCN), Chemin des \'Etoiles~8, B-1348 Louvain-la-Neuve, Belgium}
\date{\today}

\begin{abstract}
In order to make accurate predictions of material properties, current machine-learning approaches generally require large amounts of data, which are often not available in practice. In this work, an all-round framework is presented which relies on a feedforward neural network, the selection of physically-meaningful features and, when applicable, joint-learning. Next to being faster in terms of training time, this approach is shown to outperform current graph-network models on small datasets. In particular, the vibrational entropy at 305~K of crystals is predicted with a mean absolute test error of 0.009 meV/K/atom (four times lower than previous studies). 
Furthermore, joint-learning reduces the test error compared to single-target learning and enables the prediction of multiple properties at once, such as temperature functions.
Finally, the selection algorithm highlights the most important features and thus helps understanding the underlying physics.
\end{abstract}

\maketitle
\section{Introduction}
Designing new high-performance materials is a key factor for the success of many technological applications~\cite{mageeQuantificationRoleMaterials2012}.
In this respect, Machine Learning (ML) has recently emerged as a particularly useful technique in materials science (for a review, see e.g. Refs.~\cite{butlerMachineLearningMolecular2018, schmidtRecentAdvancesApplications2019,nohMachineenabledInverseDesign2020}).

Complex properties can indeed be predicted by surrogate models in a fraction of time with almost the same accuracy as conventional quantum methods, allowing for a much faster screening of materials.

Many studies have been published lately, differing by the feature generation approaches or the underlying ML models.
Concerning crystalline solids, the majority of methods presented up to now can mainly be divided in three categories. The first one, called 'ad hoc' models here, relies on a case per case study, targeted on a specific group of materials and a specific property. Typically, hand-crafted descriptors are tailored in order to suit the physics of the underlying property and are the major point of attention, while common simple-to-use ML models are chosen. Some examples include the identification of Heusler compounds of type $AB_2C$~\cite{oliynykHighThroughputMachineLearningDrivenSynthesis2016}, force field fitting by using many-body symmetry functions~\cite{behlerAtomcenteredSymmetryFunctions2011}, the prediction of magnetic moment for lanthanide-transition metal alloys~\cite{lamphamMachineLearningReveals2017} or formation energies by the sine-coulomb-matrix~\cite{faberCrystalStructureRepresentations2015}. This type of method is popular because it is simpler to construct case by case descriptors, motivated by intuition, than general all-round features. Furthermore, by focusing on a specific problem, good accuracy is often achieved. For instance, performance is increased when learning on a particular structure, which is therefore inherently built into the model.

The second category, that appeared more recently, gathers more general models that are applicable to various materials and properties based on graph networks. They transform the raw crystal input into a graph and process it through a series of convolutional layers, inspired by deep learning as used in the image-recognition field~\cite{lecunDeepLearning2015a}.
Examples of such graph models are the Crystal Graph Convolutional Neural Network (CGCNN)~\cite{xieCrystalGraphConvolutional2018} or the MatErials Graph Network (MEGNet)~\cite{chenGraphNetworksUniversal2019}.

Graph models are very convenient as they can be used for any material property. However, their accuracy crucially depends on the quantity of the available data. Since the problems that would benefit the most from machine learning are the ones that are computationally demanding with conventional quantum methods, they are precisely those for which less data is available. For instance, the band gap has been computed within $GW$ for 80 crystals~\cite{vansettenAutomationMethodologiesLargescale2017}, the lattice thermal conductivity for 101 compounds~\cite{sekoPredictionLowThermalConductivityCompounds2015}, and the vibrational properties for 1245 materials~\cite{petrettoHighthroughputDensityfunctionalPerturbation2018}). It is therefore important to develop techniques that can deal efficiently with limited datasets.
This has resulted in a third category of models trying to bridge the gap between the two former ones and combining their advantages. Examples are the sure independence screening and sparsifying operator (SISSO)~\cite{ouyangSISSOCompressedsensingMethod2018}, Automatminer~\cite{dunnBenchmarkingMaterialsProperty2020}, CrabNet~\cite{wangCompositionallyRestrictedAttentionBasedNetwork2020a} and AtomSet~\cite{chenAtomSetsHierarchicalTransfer2021}.

The present article introduces a model that falls in this third category. 
It is based on three key aspects for achieving good performance on small datasets: physically meaningful features, feature selection, and joint-learning.
We show that this framework is very effective in predicting various properties of solids with small datasets and why feature selection is important in this regime.
Finally, the selection algorithm also allows one to identify the most important features and thus helps understanding the underlying physics.

\section{Results}

\subsection{The MODNet model}

\begin{figure}[!htb]
\includegraphics{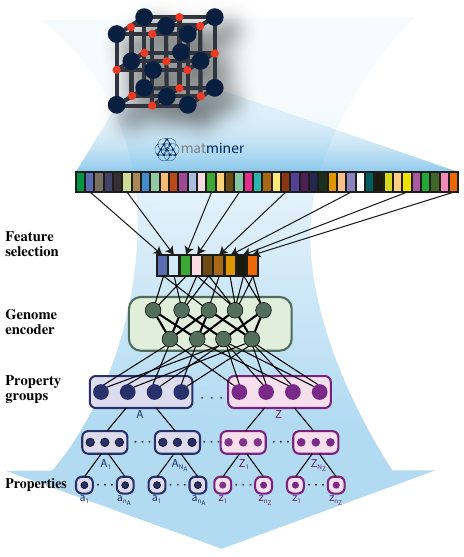}
\caption{
\textbf{Schematic of the MODNet model.}
The feature selection on matminer is followed by a hierarchical tree-like neural network. Various properties $A_1$,\ldots,$A_{N_A}$,\ldots,$Z_1$,\ldots,$Z_{N_Z}$ (e.g. Young's modulus, refractive index, ...) are gathered in groups from $A$ to $Z$ of similar nature (e.g. mechanical, optical, ...). Each of these may depend on a parameter (e.g. temperature, pressure, ...): $A(a)$,\ldots,$Z(z)$. The properties are available for various values of the parameters $a_1$,\ldots,$a_{n_A}$,\ldots,$z_1$,\ldots,$z_{n_Z}$. The
first green block of the neural network encodes a material in an appropriate all-round vector, while subsequent blocks decode and re-encode this representation in a more target specific nature.
\label{fig:tree_NN}}
\end{figure}

The model proposed here consists in building a feedforward neural network with an optimal set of descriptors. This reduces the optimization space without relying on a massive amount of data. Prior physical knowledge and constraints are taken into account by adopting physically-meaningful features selected by a relevance-redundancy algorithm. Moreover, we propose an architecture that, if desired, learns on multiple properties, with good accuracy. This makes it easy to predict more complex objects such as temperature-, pressure-, or energy-dependent functions (such as the density of states). The model, illustrated in Figure~\ref{fig:tree_NN}, is thus referred to as Material Optimal Descriptor Network (MODNet). Both ideas, feature selection and the joint-learning architecture, are now detailed further.

The raw structure is first transformed into a machine-understandable representation. The latter should fulfill a number of constraints such as rotational, translational and permutational invariances and should also be unique. In this study, the structure will be represented by a list of descriptors based on physical, chemical, and geometrical properties. In contrast with more flexible graph representations, these features contain pre-processed knowledge driven by physical and chemical intuition. Their unknown connection to the target can thus be found more directly by the machine, which is key when dealing with limited datasets. In comparison, general graph-based frameworks could certainly learn these physical and chemical representations automatically but this would require much larger amounts of data, which are often not available. In other words, part of the learning is already done before training the neural network.
To do so, we rely on a large amount of features previously published in the literature, that were centralized into the matminer project~\cite{wardMatminerOpenSource2018}. These features cover a large spectrum of physical, chemical, and geometrical properties, such as elemental (e.g. atomic mass or electronegativity), structural (e.g. space group) and site-related (i.e local environments) features. We believe that they are diverse and descriptive enough to predict any property with excellent accuracy. Importantly, a subset of relevant features is then selected, in order to reduce redundancy and therefore limit the curse of dimensionality~\cite{verleysenCurseDimensionalityData2005}, a phenomenon that inhibits generalization accuracy. In particular, previous works showed the benefit of feature selection when learning on material properties~\cite{ghiringhelliBigDataMaterials2015a,ouyangSISSOCompressedsensingMethod2018}.

We propose a feature selection process based on the \emph{Normalized Mutual Information} (NMI) defined as,
\begin{equation}
\label{eq:NMI}
    \text{NMI}(X,Y) = \frac{\text{MI}(X,Y)}{(\text{H}(X)+\text{H}(Y))/2}
\end{equation}
with MI the {mutual information}, computed as described in Ref.~\cite{kraskovEstimatingMutualInformation2004} and H the \emph{information entropy} ($\text{H}(X)=\text{MI}(X,X)$). The NMI, which is bounded between 0 and 1, provides a measure of any relation between two random variables X and Y. It goes beyond the Pearson correlation, which is parametric (it makes the hypothesis of a linear model) and very sensitive to outliers.

Given a set of features $\mathcal{F}$, the selection process for extracting the subset $\mathcal{F}_S$ goes as follows. When the latter is empty, the first chosen feature will be the one having the highest NMI with the target variable $y$. Once $\mathcal{F}_S$ is non-empty, the next chosen feature $f$ is selected as having the highest relevance and redundancy (RR) score:
\begin{equation}
    \text{RR}(f) = \frac{\text{NMI}(f,y)}{\Big[\max_{f_s \in \mathcal{F}_S}\big(\text{NMI}(f,f_s)\big)\Big]^{p}+c}
    \label{eq:RR}
\end{equation}
where $(p,c)$ are two hyperparameters determining the balance between relevance and redundancy. In practice, varying these two parameters dynamically seems to work better, as redundancy is a bigger issue with a small amount of features. Practically, after some empirical testing, we decided to set $p=\max[0.1,4.5-n^{0.4}]$ and $c = 10^{-6}n^3$ when $\mathcal{F}_S$ includes $n$ features, but other functions might even work better.
The selection proceeds until the number of features reaches a threshold which can be fixed arbitrarily or, better, optimized such that the model error is minimized. When dealing with multiple properties, the union of relevant features over all targets is taken. Our selection process is in principle very similar to the mRMR-algorithm~\cite{hanchuanpengFeatureSelectionBased2005}, but it goes beyond by combining both redundancy and relevance in a more flexible way by introducing the parameters $p$ and $c$. Furthermore, it is less computationally expensive than the Correlation-based Feature Selection (CFS)~\cite{mangalComparativeStudyFeature2018} and provides a global ranking.

In contrast with what is usually done, we take advantage of learning on multiple properties simultaneously, as recently proposed for SISSO~\cite{ouyangSimultaneousLearningSeveral2019}.
This could be used, for instance, to predict temperature-curves for a particular property.

In order to do so, we use the architecture presented in Figure~\ref{fig:tree_NN}. Here, the neural network consists of successive blocks (each composed of a succession of fully connected and batch normalization layers) that split on the different properties depending on their similarity, in a tree-like architecture. The successive layers decode and encode the representation from general (genome encoder) to very specific (individual properties). Layers closer to the input are shared by more properties and are thus optimized on a larger set of samples, imitating a virtually larger dataset. These first layers gather knowledge from multiple properties, known as joint-transfer learning~\cite{liLearningForgetting2018}. This limits overfitting and slightly improves accuracy compared to single target prediction.

Taking vibrational properties as an example, the first-level block converts the features in a condensed all-round vector representing the material. Then, a second-level block transforms this representation into a more specific thermodynamic representation that is shared by many third-level predictor blocks, predicting different thermodynamic properties (specific heat, entropy, enthalpy, energy at various temperatures). A fourth-level block splits different predictors based on the actual property, but sharing different temperature predictors. Optionally, another second-level block could be built shared by mechanical third-level predictors.

\subsection{Performance assessment}
To investigate the predictive performance of MODNet, two case studies are considered for properties originating from the Materials Project (MP)~\cite{ jainMaterialsProjectMaterials2013, ongPythonMaterialsGenomics2013, ongMaterialsApplicationProgramming2015, petrettoHighthroughputDensityfunctionalPerturbation2018, naccaratoSearchingMaterialsHigh2019}.
First, we focus on single-property learning. We benchmark MODNet against MEGNet, a deep-graph model, and SISSO, a compressed-sensing method, for the prediction of the formation energy, the band gap, and the refractive index.
Second, we also consider multi-property learning with MODNet for the vibrational energy, enthalpy, entropy, and specific heat at 40 different temperatures as well as the formation energy, as the latter was found to be beneficial to the overall performance. Since some models only predict one property at a time, we compare their accuracy with that of MODNet on the vibrational entropy at 305K.
Details about the datasets, training, validation, and testing procedures are provided in the Methodology section.

Table~\ref{tab:bench_results} summarizes the results for single-property learning on a left-out test set for the formation energy, the band gap, and the refractive index.
The complete datasets for the formation energy and the band gap include 60\,000 training samples. For the band gaps, a training set restricted to the 36\,720 materials with a non-zero band gap (labeled by a superscript $nz$ in the Table) is also considered as it was done in the original MEGNet paper~\cite{chenGraphNetworksUniversal2019}. For the refractive index, the complete dataset is much more limited containing 3\,240 compounds.
In addition to these complete datasets, subsets of 550 random samples are also considered in order to simulate small datasets. The results are systematically compared with those obtained from the MEGNet and SISSO regression. Two variants of MEGNet are used: (i) with all weights randomly initialized and (ii) by fixing the first layers to the one learned from the formation energy (i.e using transfer learning as recommended by the authors when training on small datasets). 

\begin{table}[h]
\caption{\label{tab:bench_results}
\textbf{Test accuracies with small and large training size for different machine learning algorithms.}
Comparison of the mean absolute error (MAE) on a test set in the formation energy ($E_f$ in eV/atom), the band gap ($E_g$ in eV, the superscript $nz$ refers to datasets restricted to non-zero band gaps), the refractive index ($n$) for MODNet, two variants of MEGNet and SISSO as a function of the training-set size ($N_\text{train}$).
The MEGNet variant including transfer learning is indicated by a star.}
\begin{ruledtabular}

\begin{tabular}{crcccc}
\textrm{Property}&
\textrm{$N_\text{train}$}&
\textrm{MODNet}&
\textrm{MEGNet}&
\textrm{MEGNet$^*$}&
\textrm{SISSO}\\
\colrule
$E_f$\phantom{$^\text{z}$}
&    504 & \underline{0.210} & 0.342 & 0.262 & 0.312\\
$E_f$\phantom{$^\text{z}$}
&60\,000 &0.044 & \underline{0.028} & \underline{0.028} & 0.299\\

$E_g$\phantom{$^\text{z}$}
&    504 & \underline{0.71}\0 & 0.94\0 & 0.83\0 & 0.80\0 \\
$E_g$\phantom{$^\text{z}$}
&60\,000 & 0.34\0 & 0.30\0 & \underline{0.27}\0 & 0.75\0\\

$E_g^\text{nz}$
&    504 & \underline{0.87}\0 & 0.98\0 & 0.96\0 & 0.94\0\\
$E_g^\text{nz}$
&36\,720 & 0.45\0 & 0.38\0 & \underline{0.33}\0 & 0.86\0\\

$n$\phantom{$_g^\text{nz}$}
& 3\,240 & \underline{0.05}\0 & 0.08\0 & 0.06\0 & 0.12\0\\
\end{tabular}
\end{ruledtabular}
\end{table}

MODNet systematically outperforms MEGNet and SISSO when the number of training samples is small, typically below $\sim$4\,000 samples, even when using transfer learning. In contrast, for the large datasets containing the formation energy and the band gap, MEGNet (even without transfer learning) leads to the lowest prediction error. SISSO was found to systematically result in higher errors, and  does not show significant improvement when increasing the training size.

Depending on the amount of available data, a clear distinction should thus be made between feature- and graph-based models. The former should be preferred for small to medium datasets, while the latter should be left for large datasets, as it will be confirmed for the vibrational properties.

For the second case study, i.e. multi-target learning,  the dataset only includes 1\,245 materials for which the vibrational properties have been computed~\cite{petrettoHighthroughputDensityfunctionalPerturbation2018}. 

\begin{figure}[ht!]
\includegraphics{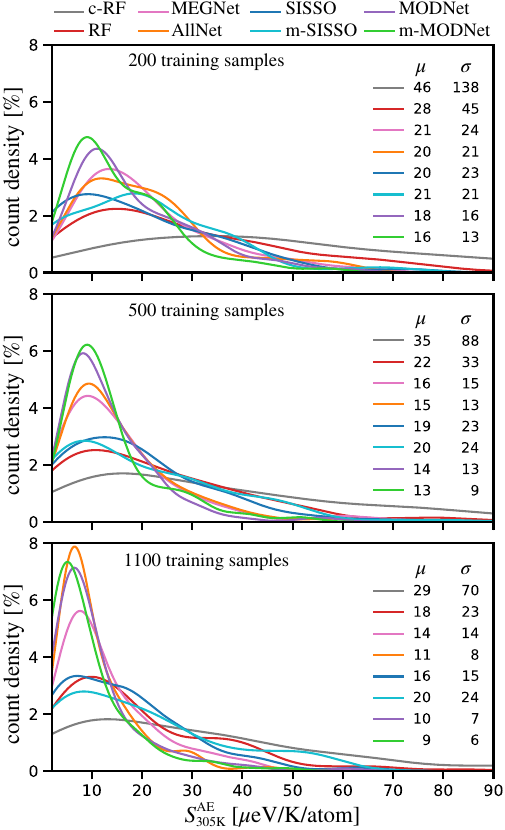}
\caption{
\textbf{Comparison of the test error distributions on the vibrational entropy for different models.}
Absolute error distribution on the vibrational entropy at 305K ($S_{305K}$ in $\mu$eV/K/atom) at three training sizes and for various strategies (see text for a detailed description). The density is obtained from a kernel density estimation with Gaussian kernel. The mean $\mu$ (equal to the MAE) and variance $\sigma$ of each distribution are also reported in $\mu$eV/K/atom.
\label{fig:samp_analysis}
}
\end{figure}

Figure~\ref{fig:samp_analysis} shows the absolute error distribution on the vibrational entropy at 305K ($S_{305K}$) at three training sizes (200, 500, and 1100 samples) for different strategies, for a systematic identical test set of 145 samples.
Furthermore, Supplementary Figure~7 reports the test MAEs as a function of the training size for the same different strategies. 
MODNet is compared with a Random Forest (RF) learned on the composition alone (i.e. a vector representing the elemental stoichiometry) similar to a previous work relying on 300 vibrational data~\cite{legrainHowChemicalComposition2017}. This strategy is referred to as c-RF in order to distinguish it from another strategy, labeled RF, which consists in a RF learned on all computed features (covering compositional and structural features). Note that, for both c-RF and RF, performing feature selection on the input space has no effect on the results as a RF intrinsically selects optimal features while learning.
This strategy can be seen as the baseline performance. The state-of-the art methods MEGNet \emph{with transfer learning} (i.e. using the embedding trained from the formation energy) and SISSO are also used in the comparison. Another strategy, labelled AllNet, is considered which consists of a single-output feedforward neural network, taking \emph{all} computed features into account. Finally, the results obtained with m-MODNet and m-SISSO, taking all thermodynamic data and formation energies, are also reported.

The lowest mean absolute error and variance is systematically found for the MODNet models, with a significant ($\sim8$\%) gain in accuracy for our joint-learning approach, more noticeable at lower training sizes.
The RF approaches are performing worst in our tests, with a large spread and maximum error, especially when considering only the composition. This is confirmed by a subsequent analysis of the features retained by the selection algorithm (see below): typically, the bond lengths are an important feature. Besides the MODNet models, AllNet, which is also based on physical descriptors,
provides a baseline to measure the gain in performance achieved thanks to feature selection.
In Figure~\ref{fig:samp_analysis} and even more clearly in Supplementary Figure~7 of the Supplementary Information, it can be seen that the usefulness of feature selection decreases with the training size. While, for 200 training samples, the gain is $\sim$12\%, it reduces to $\sim$5\% for 1000 training samples.

It is worth noting that, at the lower end of the training-set size (see 200 samples), SISSO has a comparable error with the other methods while offering a simpler analytic formula, which can be valuable. However, when increasing the training-set size, its error distribution does not seem to improve significantly in contrast with the other methods. Furthermore, contrary to m-MODNet, m-SISSO does not seem to provide any noticeable improvement with respect to SISSO in this example.

\begin{figure}[!htb]
\includegraphics[width=0.8\textwidth]{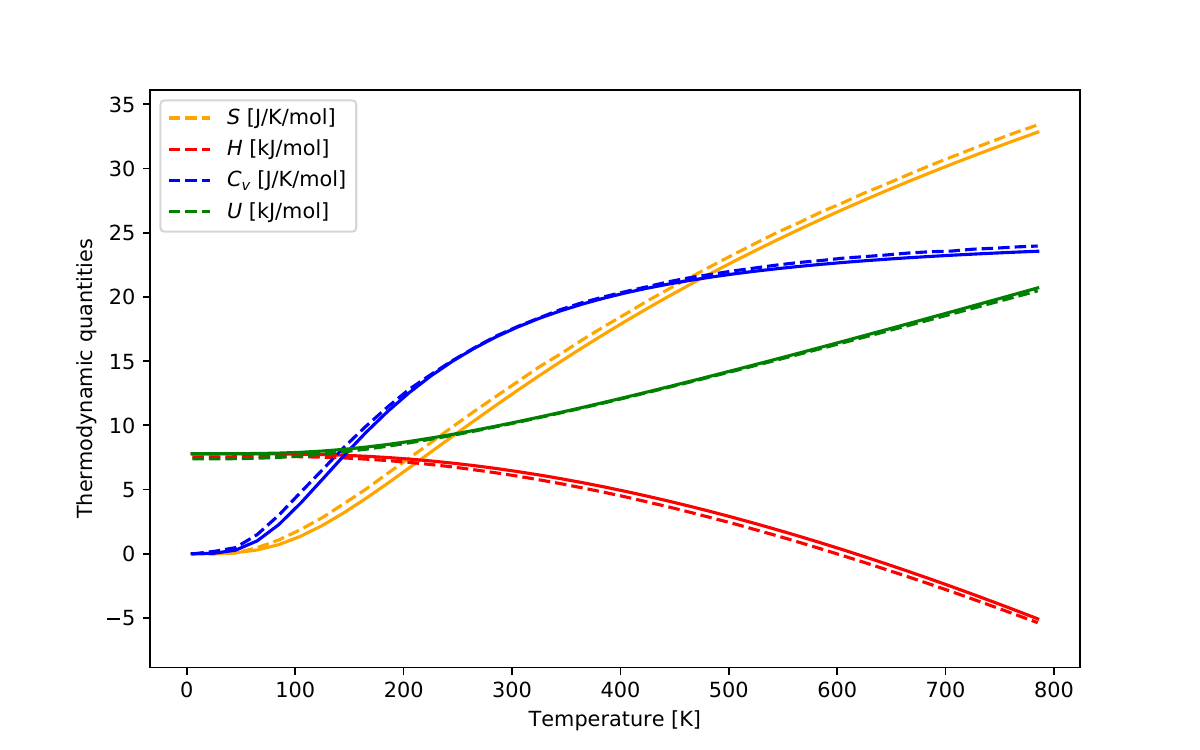}
\caption{
\textbf{Example prediction of MODNet on vibrational thermodynamics.}
MODNet predictions (dashed line) and DFPT values (solid line) for the thermodynamic quantities of Li$_2$O (MPID: mp-1960) as a function of the temperature. Observed errors on this particular sample are close to the overall MAE of the test set.\label{fig:thermo_pred_ex}}
\end{figure}

The m-MODNet was trained on four vibrational properties from 5 to 800~K: entropy, enthalpy, specific heat and Helmholtz free energy. Although the vibrational entropy at 305~K was systematically used to compare against other models, excellent performance was also found on the other properties. Table~\ref{tab:benchmark_thermo} contains the MAE for these four properties at 25, 305 and 705~K. Typical values of the corresponding properties found in the dataset are also given to compare against the error. As an example, we illustrate the prediction on Li$_2$O in Figure~\ref{fig:thermo_pred_ex}, which is a good representation of the typical observed error.

\begin{table}[htb!]
\caption{\label{tab:benchmark_thermo}%
\textbf{MODNet errors on various vibrational properties.}
MAE and MaxMAE for the vibrational entropy, Helmholtz energy, specific heat and internal energy at different temperatures as predicted with MODNet. The MaxMAE is defined as the mean over worst 5\% predicted samples.}
\small
\begin{tabular}{ l @{\qquad} c @{\qquad} c @{\qquad} c}
\toprule
\textrm{Property}& \textrm{Typical values} & \textrm{MAE ($\times 10^{-3}$)} & \textrm{MaxMAE ($\times 10^{-3}$)} \\
\colrule
$S_{25K}$ [meV/K/atom] & $\sim 10^{-5}$ -  $0.1$ & 2.5 &  2.9\\
$S_{305K}$ [meV/K/atom] & $\sim 0.03$ - $0.7$ & 9.5 & 11.3\\
$S_{705K}$ [meV/K/atom] & $\sim 0.1$ - $0.9$ & 10.8 & 12.6\\

$H_{25K}$ [eV/atom]& $\sim 7$ - $180$& 2.6 & 2.9\\
$H_{305K}$ [eV/atom] &$\sim -130$ - $180$ & 6.9 & 7.5\\
$H_{705K}$ [eV/atom] & $\sim -460$ - $150$& 8.2 & 9.7\\

$C_{v,25K}$ [meV/K/atom] & $\sim 10^{-5}$ - $0.16$& 3.2 & 3.8\\
$C_{v,305K}$ [meV/K/atom] &$\sim 0.07$ - $0.26$ & 2.6 & 3.0\\
$C_{v,705K}$ [meV/K/atom] &$\sim 0.18$ - $0.26$ & 1.6 & 1.9\\

$U_{25K}$ [eV/atom] &$\sim 10$ - $180$ & 2.2 & 2.6\\
$U_{305K}$ [eV/atom] &$\sim 80$ - $200$ & 1.6 & 1.6\\
$U_{705K}$ [eV/atom] &$\sim 180$ - $270$ & 1.2 & 1.4\\

\botrule
\end{tabular}
\end{table}

We want to emphasize that the gain in accuracy provided by joint-learning is strongly influenced by the architecture choice. The similarity between target properties is used to decide where the tree splits, i.e., the layer up to which properties share an internal representation. In all generality, one can count the number of neurons and layers that separates two properties. This determines to which degree those two properties are related. Increasing this distance (i.e. more layers and neurons between them) gives more freedom to the weights and improves learning of dissimilar properties. However, increasing it too much will tend to make the predictions independent, and no common hidden representation can be used to improve generalization. A good balance thus needs to be found between freedom and generalization. Note that increasing the architecture-distance between two properties will always decrease training error (up to convergence), but the validation error will have a minimum. Unfortunately finding this minimum based on a quantitative analysis of the dataset is rarely feasible, similarly as finding the right architecture a priori for a single-target model. It is therefore considered as a hyperparameter, as it is commonly done in the ML field. In practice, we suggest to first gather the properties in groups and subgroups based on their similarity. This will define the splits in the tree-like architecture. Then, various sizes for the layers and number of neurons (which will define the intra-property distance in architectural space) should be included in the regular hyperparameter optimization of the model. An in-depth example for the architectural choice for the vibrational properties can be found in the Supplementary Information, section C.

\subsection{Feature selection}
Feature selection is a valuable asset of MODNet and has two main advantages. First, it was shown in Figure~\ref{fig:samp_analysis} that an average 12\% improvement in error can be obtained by removing irrelevant features. This is far from negligible. This increase in performance is achieved by reducing the noise to signal ratio, caused by the curse of dimensionality. This is especially the case for small datasets. Supplementary Figure~7 shows that the gain in performance by feature selection reduces as the training size increases. We therefore expect that feature selection will be less important for larger datasets.

\begin{figure}[!htb]
\includegraphics[width=0.9\textwidth]{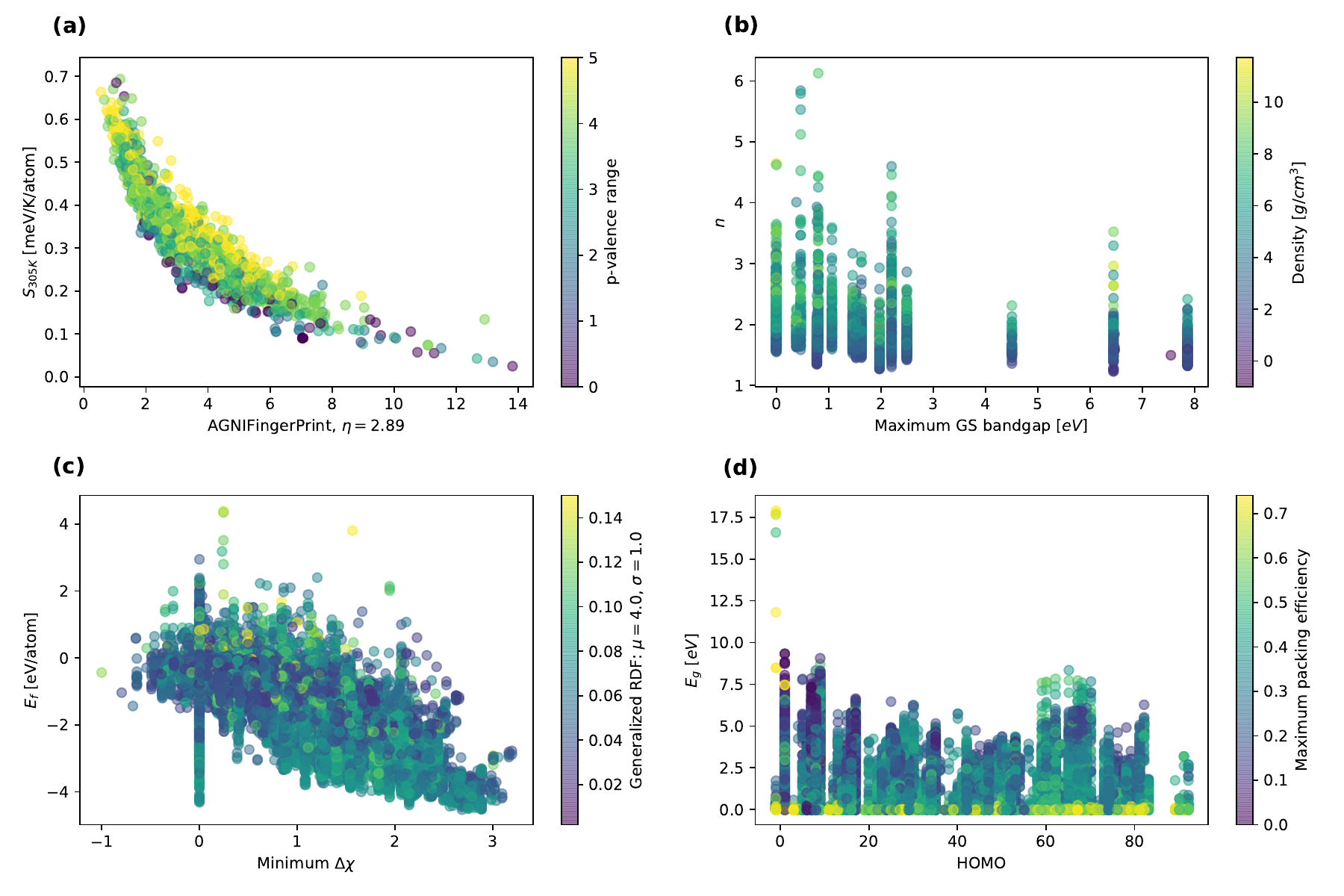}
\caption{
\textbf{Visualization of selected features.}
Bivariate representation of the two most important features for four different properties: (a) vibrational entropy at 305~K, (b) refractive index, (c) formation energy and (d) band gap energy. Both features are complementary to narrow down the target output, although certainly not sufficient for an accurate estimation. \label{fig:selection_vis}}
\end{figure}

Second, feature selection (compared to feature extraction) has the advantage of keeping the input space understandable. As they are chosen according to their relation (i.e. mutual information), important factors contributing to the target property can be found. Figure~\ref{fig:selection_vis} shows a bivariate visualization for the vibrational entropy, formation energy, band gap and refractive index as a function of the two first selected features. Thanks to the redundancy criterion, both features are complementary to predict the target. A detailed description of these features can be found in Sec.~B of the Supplementary Information. 
Concerning the vibrational entropy, a strong correlation is seen with the first feature, namely AGNIFingerprint, which gives a measure of the inverse bond length. In other words, increasing the average bond length increases the vibrational entropy. Similarly having a larger range of p-valence electrons (which is linked to ionicity) increases the vibrational entropy. 
Concerning the refractive index, two import factors are identified: the band gap and the density of the material. The band gap, although not explicitly given but instead approximated by the bandgap of the constituent elements, is known to be an important variable. Typically there is an inverse relation between the band gap energy and the refractive index, see Ref.~\cite{naccaratoSearchingMaterialsHigh2019}. Finding materials combining a high value for both properties remains a tedious task, and could therefore certainly benefit from Machine Learning.
Overall, it is seen how common intuitive patterns for the physicist are indeed retrieved by the machine. Therefore, this strategy can be used to analyze and find underlying factors for all types of properties and datasets.

The feature selection algorithm presented in this work is based on relevance and redundancy and will be called \textit{MOD-selection}.
Other popular choices exist.
Here, MOD-selection is compared to five other algorithms:
(i) \textit{corr-selection} in which features having the highest Pearson-correlation with the target are selected first;
(ii) \textit{NMI-selection} in which features having the highest NMI with the target are selected first;
(iii) \textit{RF-selection} where the data is first fitted with a Random Forest (300 trees) and features are ranked according to their impurity-based importance;
(iv) \textit{SISSO-selection} in which the data is first fitted by the SISSO model without applying any operator on the feature set, i.e. only primary features are used (rung set to 0) and each $n^{th}$ dimension of the final model corresponds to the n-th descriptor;
and (v) \textit{OMP-selection} in which an orthogonal matching pursuit is applied by using the SISSO strategy with a SIS-space restricted to one.

It is worth noting that, although SISSO is a powerful dimensionality reduction technique, it can not be used as such for feature selection with the same generality as the other techniques.
Indeed, SISSO provides a general framework for selecting the best few descriptors from an immense set of candidates but the selection is computationally limited to $\sim$10 features.
This is not an issue for the original aim of SISSO (which consists in a low dimensional model), but it surely is when used together with a neural network, where the optimal amount is typically a few hundreds of features.
Therefore, when going beyond the 10$^{th}$ feature, we simplified SISSO to OMP, which scales linearly with the number of features.

\begin{figure}[!htb]
\includegraphics[width=\textwidth]{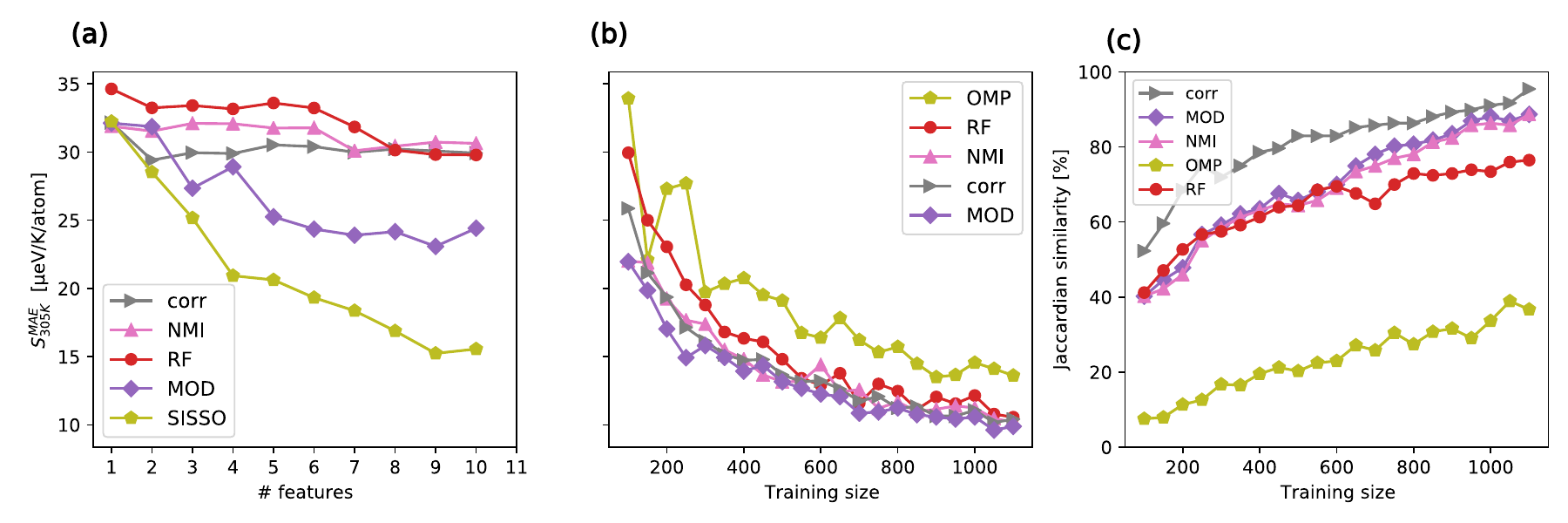}
\caption{
\textbf{Performance comparison of different feature selection methods.}
(a) Test error on the vibrational entropy at 305K for different feature selection algorithms, as a function of the first few features for 1100 training samples. (b) Test error on the vibrational entropy at 305K for different feature selection algorithms as a function of the training size with other parameters being optimized over a fixed grid. Models are constructed by replacing the selection algorithm in MODNet by a Pearson correlation (corr), Normalized Mutual Information (NMI), Random Forest (RF), SISSO, and orthogonal matching pursuit (OMP). (c) Jaccardian similarity of the 300 first selected features on a sampled training set of size n, and the total dataset (1245 samples) as a function of n, for different feature selection algorithms.
\label{fig:selection_comparison}}
\end{figure}

Figure~\ref{fig:selection_comparison}(a) shows the test error on the vibrational entropy at 305~K for the different models (MODNet substituted with different selection algorithms) for the first 10 selected features. The training size is fixed to 1100 samples. There is a clear distinction between redundancy based techniques (MOD and SISSO) and non-redundancy based techniques (corr, NMI and RF). Accounting for redundancy is clearly important when using only a few features. In this particular scenario, SISSO outperforms MOD-selection.

However, in order to construct the best possible model, one should go much beyond 10 features. Figure~\ref{fig:selection_comparison}(b) depicts the same error but as a function of the training size, all other hyperparameters being optimized. The number of optimal features is often chosen to be around 300 features. SISSO is therefore replaced by the OMP.
It can be seen how MODNet outperforms all other selection algorithms, particularly at low sample size. The OMP method performs poorly due to wrongly chosen features which result in overfitting.

As an additional experiment, we aim to measure how well a feature selection algorithm is able to capture the important features, when only given a limited amount of labeled samples. We measure this by first running each selection algorithm on the \emph{total dataset}, keeping the best 300 features, which forms the best approximation of optimal features for each algorithm. As a second step, we do the same but on sampled subsets of varying size. The Jaccardian similarity between the 300 features found on the subset and total dataset is represented in Figure~\ref{fig:selection_comparison}(c). Note that this metric only represents how fast an algorithm converges in terms of chosen features, but does not necessarily mean that the selected features are worthwile. All methods (including Pearson-correlation) suffer significantly from small datasets, with a Jaccardian-similarity change of over 40\% from 200 to 1000 samples. Similarity increases when the training size increases (with some exceptions due to sampling variance), as the sampled dataset approaches the total dataset. The correlation method provides the highest similarity for all training sizes. This can be explained by the simpler nature of the algorithm: measuring linear dependence requires less samples than more complex non-linear dependencies. The MOD and NMI approaches have a steeper increase in similarity than the RF-approach, while the three are non-linear approaches. Finally, the OMP algorithm has a low Jaccardian similarity, and this for the whole range of subsets. Additional experiments on the OMP showed that the similarity between features chosen on the different sampled subsets are also low, in contrast with the other methods. This clearly shows that a significant variance in selected features is found when slightly changing the training set, which is not desirable. Therefore the OMP method should be avoided. This also explains the poor performance in Figure~\ref{fig:selection_comparison}(b).

From these results, one can conclude that depending on the number of features to be selected, some algorithms work better than others. As soon as there are more than 10 features (which is the case for most practical problems), MOD-selection performs the best. It should however be noted that, when selecting only a few features, accounting for redundancy is critical and, in this case, SISSO was found to be best. Unfortunately, it becomes computationally unaffordable above 10 features.

\section{Discussion}

Previous results show that although state-of-the art methods such as graph networks are very powerful on big datasets, they do not scale well on smaller datasetes which are typically encountered in physics. Our framework provides excellent accuracy on limited datasets by using prior knowledge such as prepossessed meaningful features or multiple properties for a same material. Beyond increasing accuracy, the m-MODNet is also convenient for constructing a single model for multiple properties, hence speeding up training and prediction time.

We showed that feature selection is very useful for small datasets. An improvement of 12\% was found on the vibrational thermodynamics when learning on 200 samples. Moreover, an additional improvement of 8\% on $S_{305K}$ can be attributed to the joint-learning mechanism of MODNet.

Importantly, our model provides the most accurate ML-model at present for vibrational entropies with a MAE (resp. RMSE) of 8.9 (resp. 12.0) $\mu$eV/K/atom on $S_{305K}$ on a hold-out test set of 145 materials. This is four times lower than reported by Legrain~\textit{et~al.}~\cite{legrainHowChemicalComposition2017} (trained on 300 compounds) and 25 times lower than reported by Tawfik~\textit{et~al.}~\cite{tawfikPredictingThermalProperties2020} (trained on the exact same dataset as this work).

Another important advantage of MODNet is that its feature selection algorithm provides some understanding of the underlying physics. Indeed, it pinpoints the most important and complementary variables related to the investigated property. For instance, the vibrational entropy is found to strongly depend on the inter-atomic bond length and the valence range of the constituent elements (which relates to the ionicity of the bond) while the refractive index is related to an estimation of the band gap and to the density.

Although all property predictions in this work were made from structural primitives, MODNet is certainly not limited to structures. For instance, it can easily be extended to composition-only tasks (see GitHub repository~\cite{MODNetWebsite}).

In summary, we have identified a frontier between physical-feature-based methods and graph-based models. Although the latter are often referred to as state-of-the-art for many material predictions, the former are more powerful when learning on small datasets (below $\sim$4\,000 samples). We have proposed a novel model based on optimal physical features. Descriptors are selected by computing the mutual information between them and with the target property in order to maximize relevance and minimize redundancy. This combined with a feedforward neural network forms the MODNet model. Moreover, a multi-property strategy was also presented. By modifying the network in a tree-like architecture, multiple properties can be predicted, which is useful for temperature functions, with an increase in generalization performance thanks to joint-transfer learning. In particular, this strategy was applied on vibrational properties of solids, providing remarkably reliable predictions, orders of magnitude faster than conventional methods. Finally, we illustrated how the selection algorithm which determines the most important features can provide some understanding of the underlying physics.

\section{Methods}
\subsection{Datasets}
Four datasets were used throughout this work: formation energies, band gaps, refractive indices and vibrational thermodynamics.

The crystal data set for the band gaps and formation energies are based on DFT computations of 69,640 crystals from the Materials Project obtained via the Python Materials Genomics (pymatgen) interface to the Materials Application Programming Interface (API) on June 1, 2018 \cite{ongPythonMaterialsGenomics2013,ongMaterialsApplicationProgramming2015}.
Those crystals corresponds to the ones used for MEGNet (i.e. the MP-crystals-2018.6.1 dataset), which facilitates benchmarking as the Materials Project is constantly being updated. A subset of 45,901 crystals with finite band gap was used for the non-zero band gap regression (superscript \textit{nz} in Table~\ref{tab:bench_results}). 

The vibrational properties for 1,245 inorganic compounds were computed by Petretto~\textit{et al.}~\cite{petrettoHighthroughputDensityfunctionalPerturbation2018}, in the harmonic approximation based on Density Functional Perturbation Theory (DFPT). This dataset contains the following thermodynamic properties: vibrational entropy, Helmholtz free energy, internal energy and heat capacity from 5 to 800~K in steps of 5~K. Supplementary Figure~1 graphically represents these four properties from 5 to 800~K for all materials contained in the dataset in meV/atom or meV/K/atom. A wide variety of values is reached, with different Debye temperatures as can be seen from the specific heat. This indicates no significant bias, giving us confidence for generalizing on unseen data.

The refractive index for 4040 compounds were computed by Naccarato~\textit{et al.} (see Ref.~\cite{naccaratoSearchingMaterialsHigh2019}) relying on Density Functional Theory (DFT) and high-throughput methods. Typical values encountered in the dataset ranges from 1 to 6, with 60\% below 2.

Supplementary Figures~2-4 contain various histograms representing the data distribution for both latter properties. Various crystalline compounds are present ranging from simple mono-elemental compounds to complex semiconductors. The thermodynamic data (resp. refractive index) cover the 7 (resp. 7) symmetry groups, 84 (resp. 165) space groups and 64 (resp. 86) elements.
Most compounds are ternary alloys and there are no (resp. almost no) materials with more than 5 different elements. The mean atomic mass has a large range, confirming the variety of elements present in the two datasets. Note that the refractive index dataset is biased towards oxides, with 84\% of all materials containing at least one oxygen atom.

\subsection{Model training}

For assessing the performance of a model, we follow the standard procedure which consists in splitting the dataset in mutually exclusive training, validation and test sets. Validation is used in order to optimize the hyperparameters, while the test set is used for obtaining an unbiased generalization performance for the best hyperparameters. The Mean Absolute Error (MAE) is systematically used as performance criterion, except on the vibrational thermodynamics where a large set of metrics is used to fully capture the multiple properties learned at once (see the Supplementary Information for further information). Moreover, all datasets that are considered as small (i.e. all properties except the full formation and band gap energy sets covering 60\,000 training samples) use a 10-fold validation scheme. Supplementary Table~1 summarizes training, validation and test set sizes for each property used in this work. 

\begin{figure}[!htb]
\includegraphics[width=0.8\textwidth]{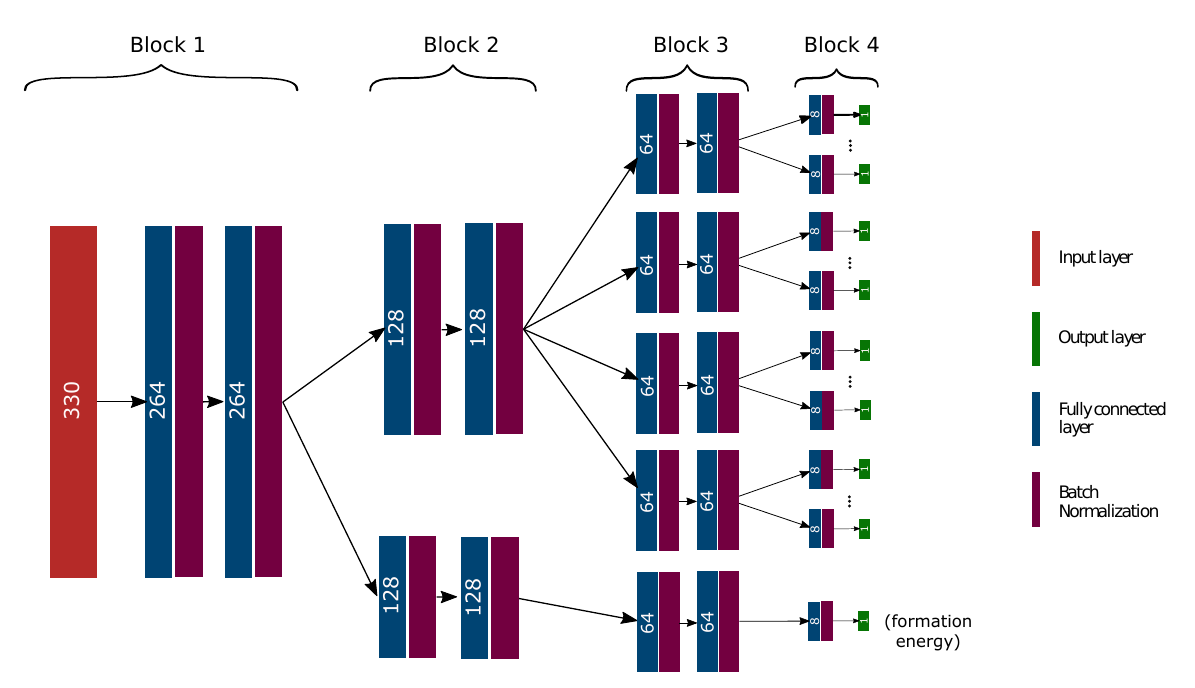}
\caption{
\textbf{MODNet architecture for the vibrational properties.}
Architecture of the MODNet (composed of 4 blocks) when learning on the vibrational properties. The formation energy is added by adding a second order block to the first block.\label{fig:architecture}}
\end{figure}

The hyperparameters that were tuned are the following. For neural network based models they consist in the number of layers, the number of neurons per layer, the learning rate, the batch size, the activation function and finally the loss function. The MODNet model has an additional hyperparameter consisting in the number of optimal input features. Similarly, the MEGNet has also an additional hyperparameter consisting in the number of MEGNet-blocks. Finally, when using the Random Forest the number of trees is taken as the only hyperparameter.
In Supplementary Information, section C, an in depth-example is given on how the hyperparameters were chosen for MODNet when trained on multiple vibrational thermodynamic quantities. The final model has a min-max preprocessing, learning rate set to 0.01, MSE loss (with scaling of targets, see Supplementary Information), an architecture of two layers per block and 256, 128, 64 and 8 neurons in these succesive blocks. Adding (or removing) a layer, as well as doubling or halving the number of neurons does not improve accuracy as can be seen in Supplementary Figure~5. The batch size was fixed to 256. A rectified linear unit function (ReLU) is used as activation for each layer. Learning is performed using an Adam optimizer ($\beta_1=0.9$, $\beta_2 = 0.999$, $decay = 0$) on 600 epochs. The final architecture is depicted in Figure~\ref{fig:architecture}.

\section{Data availability}
The generated features, NMI and MP-2018.6 datasets are available on \url{https://figshare.com/account/home#/projects/82607}. The vibrational thermodynamics and refractive index datasets are respectively available from Refs.~\cite{petrettoHighthroughputDensityfunctionalPerturbation2018} and~\cite{naccaratoSearchingMaterialsHigh2019}.

\section{Code availability}
The modnet python package with pretrained models is available as a package on GitHub from Ref.~\cite{MODNetWebsite}.

\section{acknowledgments}
The authors acknowledge useful discussions with R.~Ouyang and L.~Ghiringhelli about the SISSO framework. 
P.-P.~D.B. and G.-M.~R. are grateful to the F.R.S.-FNRS for financial support.
Computational resources have been provided by the supercomputing facilities of the Université catholique de Louvain (CISM/UCL) and the Consortium des Équipements de Calcul Intensif en Fédération Wallonie Bruxelles (CÉCI) funded by the Fond de la Recherche Scientifique de Belgique (F.R.S.-FNRS) under convention 2.5020.11 and by the Walloon Region. G.~H. acknowledges funding by the U.S. Department of Energy, Office of Science, Office of Basic Energy Sciences, Materials Sciences and Engineering Division, under Contract DE-AC02-05-CH11231: Materials Project program KC23MP.

\section{Author contributions}
P.-P.~D.B., G.~H. and G.-M.~R. conceived the project and prepared the manuscript. P.-P.~D.B designed, implemented and benchmarked the model. G.-M.~R. supervised the project.

\section{Competing Interests}
The authors declare no competing interests.

\section{Figure Legends}

Figure 1.
\textbf{Schematic of the MODNet model.}
The feature selection on matminer is followed by a hierarchical tree-like neural network. Various properties $A_1$,\ldots,$A_{N_A}$,\ldots,$Z_1$,\ldots,$Z_{N_Z}$ (e.g. Young's modulus, refractive index, ...) are gathered in groups from $A$ to $Z$ of similar nature (e.g. mechanical, optical, ...). Each of these may depend on a parameter (e.g. temperature, pressure, ...): $A(a)$,\ldots,$Z(z)$. The properties are available for various values of the parameters $a_1$,\ldots,$a_{n_A}$,\ldots,$z_1$,\ldots,$z_{n_Z}$. The
first green block of the neural network encodes a material in an appropriate all-round vector, while subsequent blocks decode and re-encode this representation in a more target specific nature.
\\

Figure 2.
\textbf{Comparison of the test error distributions on the vibrational entropy for different models.}
Absolute error distribution on the vibrational entropy at 305K ($S_{305K}$ in $\mu$eV/K/atom) at three training sizes and for various strategies (see text for a detailed description). The density is obtained from a kernel density estimation with Gaussian kernel. The mean $\mu$ (equal to the MAE) and variance $\sigma$ of each distribution are also reported in $\mu$eV/K/atom.
\\

Figure 3.
\textbf{Example prediction of MODNet on vibrational thermodynamics.}
MODNet predictions (dashed line) and DFPT values (solid line) for the thermodynamic quantities of Li$_2$O (MPID: mp-1960) as a function of the temperature. Observed errors on this particular sample are close to the overall MAE of the test set.
\\

Figure 4.
\textbf{Visualization of selected features.}
Bivariate representation of the two most important features for four different properties: (a) vibrational entropy at 305~K, (b) refractive index, (c) formation energy and (d) band gap energy. Both features are complementary to narrow down the target output, although certainly not sufficient for an accurate estimation.
\\

Figure 5.
\textbf{Performance comparison of different feature selection methods.}
(a) Test error on the vibrational entropy at 305K for different feature selection algorithms, as a function of the first few features for 1100 training samples. (b) Test error on the vibrational entropy at 305K for different feature selection algorithms as a function of the training size with other parameters being optimized over a fixed grid. Models are constructed by replacing the selection algorithm in MODNet by a Pearson correlation (corr), Normalized Mutual Information (NMI), Random Forest (RF), SISSO, and orthogonal matching pursuit (OMP). (c) Jaccardian similarity of the 300 first selected features on a sampled training set of size n, and the total dataset (1245 samples) as a function of n, for different feature selection algorithms.
\\

Figure 6.
\textbf{MODNet architecture for the vibrational properties.}
Architecture of the MODNet (composed of 4 blocks) when learning on the vibrational properties. The formation energy is added by adding a second order block to the first block.
\\
\end{document}